# Optical multi-beam steering and communication using integrated acousto-optics arrays

Qixuan Lin[1], Shucheng Fang[1], Yue Yu[1], Zichen Xi[3], Linbo Shao[3,4], Bingzhao Li[1*], and Mo Li[1,2†]

[1]Department of Electrical and Computer Engineering, University of Washington, Seattle, WA, USA
[2]Department of Physics, University of Washington, Seattle, WA, USA
[3]Bradley Department of Electrical and Computer Engineering, Virginia Tech, Blacksburg, VA, USA
[4]Center for Quantum Information Science and Engineering (VTQ), Virginia Tech, Blacksburg, VA, USA

**Optical beam steering enables optical detection and imaging in macroscopic or microscopic scales and long-range communication over free space. It underpins numerous optical applications, including LiDAR, biomedical imaging, and remote sensing. Despite the inherent speed of light, advanced applications increasingly require the ability to steer multiple beams simultaneously to increase imaging throughput, boost communication bandwidth, and control arrays qubits for scalable quantum computing. Therefore, there is a significant demand for non-mechanical, integrated, and scalable multi-beam steering technology. Here, we report a scalable multi-beam steering system comprising an array of acousto-optic beam steering channels and photonic integrated circuits on a thin-film lithium niobate platform. Each channel generates tens of individually controllable beams of visible wavelength by exciting acoustic waves using digitally synthesized multi-tone microwave signals. We demonstrate the system's capabilities through multi-input, multi-output free-space communications, simultaneously transmitting to multiple receivers at megabits/sec data rates. This technology can be readily scaled up to steer hundreds of optical beams from a compact chip, potentially advancing many areas of optical technologies and enabling novel applications.**

---

[*] Email:bzli@uw.edu
[†] Email: moli96@uw.edu

**Main**

Solid-state optical beam steering is crucial for a wide array of optical technologies (Fig. 1a), including light detection and ranging (LiDAR)[1,2], bioimaging[3,4], optical trapping[5–7], and performing parallel gate operations of optically addressed qubits, such as neutral atoms[8–11], trapped ions[12–14], and defect centers in solids[15,16]. It also enables free-space optical communications (FSOC)[17–19]. Various solid-state beam steering technologies have been developed, including optical phased arrays (OPA)[20–22], focal plane switch arrays (FPSA)[23,24], digital micromirror devices (DMD)[25,26], and spatial light modulators (SLM)[27]. These technologies share a common principle: they steer light by creating a synthetic aperture composed of an array of microscale pixels, which relies on sophisticated peripheral control circuits to precisely control optical amplitude or phase at each pixel. Demands on wider steering angles and higher steering resolutions necessitate an increasing number of pixels, presenting significant scaling challenges. While existing solid-state beam steering technologies primarily focus on steering a single beam, multi-beam steering[28,29] could address performance requirements in advanced applications. For instance, the imaging frame rate of a single-beam-scanning LiDAR system is fundamentally limited by the time of flight and insufficient for autonomous driving systems. This limitation can be addressed by scanning multiple beams simultaneously[28,29]. Multi-beam steering also enables scalable quantum computing using optically addressed qubits[11]. In these application scenarios, performance metrics such as beam quality, overall scanning speed, and the ability to independently control individual beams are important.

Achieving multi-beam steering with SLM or OPA is possible, but they face challenges that make them impractical for many applications. For example, SLMs can segment their aperture into several sub-apertures to diffract and scan multiple beams. However, the scanning speed is limited to the kHz range, which is insufficient for high-speed applications[27]. OPAs can generate multiple beams by superimposing multiple phase patterns on the array, but they suffer from low extinction ratios due to high crosstalk between the beams[30,31]. In general, multi-beam steering using integrated photonic systems must overcome several common challenges. The first is power handling capability: each beam must carry the same optical power as in single-beam mode, significantly increasing the total power handling requirement. The second challenge is the increased crosstalk between different beams, which is caused by reduced beam quality when multiple beams are generated from the same aperture. Lastly, there is the challenge of controlling and modulating individual beams so that they can be utilized independently and multiplexed to enhance sensing and imaging throughput or overall system communication bandwidth. The limitations of existing technology call for the innovation of new optical beam steering techniques.

The emerging integrated acousto-optic devices could provide a powerful solution to overcome the abovementioned challenges. Acousto-optic devices utilize acoustic waves that are electromechanically excited in materials to generate a moving phase mask, which coherently modulates and shapes optical waves that pass through. Conventional acousto-optic devices using bulk materials, including acousto-optic deflectors (AODs) and modulators (AOMs), have already proven to be effective as optical switches and intensity modulators in various laser tools[32,33]. Recently, AODs operated in the multi-beam mode have played a critical role in performing parallel gate operations on neutral atom arrays for quantum computing[8–11]. With the emergence of new material platforms and the advancement of optomechanical systems, acousto-optics have been incorporated into integrated photonics, demonstrating versatile applications on various platforms[34–38]. Previously, on a lithium niobate on insulator (LNOI) platform, we demonstrated an integrated acousto-optic beam steering (AOBS) device in which GHz frequency acoustic waves are generated to steer waveguided light into free space with a wide field of view and diffraction-limited resolution[39]. In contrast to schemes using pixelated apertures, AOBS has a continuous aperture and uses only a single acoustic transducer to control beam steering. Moreover, the coherent acousto-optic scattering process allows one AOBS device to be driven simultaneously by multiple control signals of different frequencies. Each tone of the generated acoustic wave scatters light into a different angle to achieve multi-beam steering. The compact size of AOBS allows multiple devices to be closely integrated on one chip, proportionally increasing the number of steerable beams. Furthermore, the amplitude and phase of each beam can be independently modulated by the corresponding control signal with an overall bandwidth of hundreds of MHz.

In this work, we report an integrated multi-beam AOBS (mAOBS) array and demonstrate free space optical communication (FSOC) with multiple-input multiple-output (MIMO) by combining multi-beam steering and modulation. The mAOBS features an array of monolithically integrated channels, each capable of generating and steering more than twenty beams with high beam quality and superior dynamic performance. By modulating each beam using Pulse Amplitude Modulation (PAM) or Quadrature Amplitude Modulation (QAM) coding, we achieve MIMO communication with an aggregate bandwidth exceeding 100 megabits per second (Mbps). MIMO-enabled FSOC affords precise and swift multi-channel communication between decentralized drone swarms or autonomous vehicles, ensuring secure and efficient coordination[39]. It also enables fast-deployed, reconfigurable, and broad bandwidth satellite or air-to-ground communications[41]. Our results demonstrate that the mAOBS array is a scalable and flexible multi-beam steering solution that has the potential to enable many emerging optical technologies (Fig. 1a).

**Results**

**Multi-beam AOBS Array**

Our mAOBS array consists of three main components (Figs. 1b and 1c): interdigital transducer (IDT) arrays for surface acoustic wave (SAW) generation; a continuous acousto-optic (AO) aperture for beam steering, and photonic integrated circuits (PICs) for on-chip light coupling, distributing, and shaping. The AO aperture comprises multiple AOBS channels that are formed by collimated acoustic and optical beams, rather than by patterning, to maximize the interaction area and reduce background scattering by etched structures. In each channel, the SAW is launched by the IDT to scatter light off the chip and steer along the horizontal axis (H-axis). The array of AOBS channels is distributed along the vertical axis (V-axis), as denoted in Fig. 1b. We place the mAOBS chip at the focal plane of a cylindrical convex lens, which collimates the beams from different channels. With each AOBS channel driven by multiple control signals of different frequencies, a 2D array of beams is generated. In addition, each beam can be individually modulated by the corresponding control signal at each channel.

The device is fabricated on an X-cut lithium niobate on insulator (LNOI) substrate with 300 nm thick LN layer. A laser with 780 nm wavelength (Newport TLB-6712) is coupled from a fiber to the chip through a waveguide edge coupler. PICs (Fig. 1f) of single mode waveguides are patterned and etched in the LN layer to route and distribute light to different channels on the chip. Before entering the AO aperture, the waveguide mode is expanded to the 30-$\mu$m-wide slab mode to increase the AO interaction area and aperture width. At the end of the mode expander, the slab mode is converted from the $TE_{00}$ mode to the $TE_{01}$ mode by a mode converter (Fig. 1e), and then freely propagates into the AO aperture. Our simulation has shown that the $TE_{01}$ mode can be more efficiently steered by the SAW, propagating along Y-axis of the LN layer, than the $TE_{00}$ mode due to the enhanced moving boundary effect (See Supplementary Information). The pitch ($d$) between the AOBS channels is 250 $\mu$m, sufficient to prevent crosstalk. On the other end of the AO aperture, an array of broadband IDTs is fabricated. The IDT periods are linearly chirped from 1.05 to 1.25 μm to generate SAW with frequencies centered at $f_0$=2.0 GHz over a bandwidth of $BW$=450 MHz. Each IDT is wire-bonded to a microwave circuit board to be connected to signal sources.

**Multi-Beam Steering**

We characterized the mAOBS array's steering performance by imaging the 2D beam array using a charge-coupled device (CCD) camera placed on the image plane of the cylindrical lens. Four AOBS channels are operated simultaneously. Fig. 2a shows a superimposed image of an array of 4×21 beams captured by the camera. The beams are generated by scanning the control signal frequency across the IDT bandwidth with a step of $\Delta f$ =15 MHz. Each row of the beams is generated by each AOBS channel. The distortion of the beam in the second row is attributed to a small blemish of contamination on the surface of the chip. The average beam divergence is 0.19°. By fitting the beam's intensity profile, we determine the effective aperture length ($l$) of the AOBS channel is 373 μm, limited by acoustic loss. The best beam steering efficiency of a single beam is 3.3% (-14.8 dB), referred to the optical power in the waveguide, when 14.8 dBm control signal power is used (See Supplementary Information). The theoretical number of resolvable spots of a single channel is given by $N_1 = BW \cdot l/v = 54$, using the SAW velocity of $v$=3,110 m/s. For multiple channels, each controlled by different control signals, the total number of resolvable spots $N_t$ simply increases proportionally with channel count. The area-normalized resolvable spot density of our device thus is $\sigma = N_t/(d \cdot l)$ =579 (spots/mm$^2$), where $d$ is the width of the whole AO aperture. Compared with conventional bulk AODs (ISOMET D55-T80S-2, $\sigma$=64 spots/mm$^2$), the mAOBS array has a significant advantage.

Another important metric of beam steering systems is the extinction ratio between beams. Fig. 2b shows beam profiles of 4, 8, and 16 beams generated by a single AOBS channel (cyan lines). Because the extinction ratio exceeds the dynamic range of the CCD camera (14 dB), to quantify the contrast and crosstalk between beams, we collected the steered beams by scanning a single-mode fiber on the focal plane of the collimating lens and measured the optical power with a photodetector. For comparison, the background signal, when the SAW is turned off, is plotted as the shaded region. We used an arbitrary waveform generator (Tektronix AWG70001A) to synthesize control signals with up to 16 frequency tones in the range from 2.35 to 2.65 GHz with $\Delta f$ =20 MHz. The corresponding angular positions of the 16 beams are marked with circles in

Fig. 2b. The x-axis of Fig. 2b is the beam position on the image plane, normalized to the average beam width $w_0$=43.9 μm. When four beams are generated simultaneously using four frequency tones with a total power of $P_T$=16 dBm, the maximum (average) on-off contrast achieved is 29.7 (27.8) dB. As more beams are generated, the on/off contrast gradually reduces because of reduced power at each RF tone when $P_T$ is fixed. The IDT has a limited power handling capability, beyond which it can be damaged due to overheating in ambient conditions. This power limit can be improved using more robust material with protection by a passivation layer and by operating in vacuum or inert gas environment. As shown in Fig. 2b, the maximum (average) on/off contrast is 27.4 (24.8) dB for 8 beams and 27.8 (20.9) dB for 16 beams. While the beam power is uniform with less than 1% variation, the background has a larger variation due to scattering by etched device structures on the chip. From the recent results of high-Q LN ring resonators[42–44], we can estimate the optical scattering limit of LN photonic structures and expect the contrast to be further improved to above 60 dB. The crosstalk between beams separated by $5w_0$ (~1° in angular position) is ~-20 dB, reducing to -25 dB at $10w_0$ separation. Fig. 2c shows the detailed 2D mapping of 16 concurrently generated beams, featuring uniform beam shapes and contrasts across the whole field of view.

**Beam steering dynamics and communication**

Besides the multi-beam steering capability, the mAOBS array also shows superior dynamic performance than conventional technologies. We first characterize the linearity, accuracy, and stability of the beam power controlled by the control signal. Fig. 3a plots the measured power of a steered beam versus the control signal power varied from 0 to 13 dBm. The optical power is measured with a photodiode using 100 $\mu$s integration time. We repeated the measurement at each control signal power level for over $3\times10^4$ switching cycles. The result and linear fitting of it in Fig. 3a show excellent agreement, and the normalized standard deviation (SD) is very low at 0.42%. Fig. 3b displays the photodiode voltage waveform captured on an oscilloscope when the control signal is switched on, showing a very short switching rising time of 116 ns (10% to 90%).

  The demonstrated linearity, repeatability, and speed of AOBS are important attributes for applications such as optical control of qubits and FSOC. To demonstrate the latter, we modulated the control signal and encode pseudo-random pulse sequence data using various coding schemes. The data was carried by the steered beams and transmitted over free space in a direction controlled by the frequency of the control signal. The steered beam was detected by a receiver and characterized by performing eye diagram and constellation diagram analysis. Fig. 3c shows measured eye diagrams when on-off keying (OOK) and 4-level pulse amplitude modulation (PAM-4) coding are used, respectively. The data rate is 4 Mbps, which is equivalent to Bluetooth 5.0 transmission rate. Both diagrams show clear eye openings, with a quality factor of 14, indicating a high signal-to-noise ratio. We then advanced the coding scheme to 64-level quadrature amplitude modulation (64-QAM). Fig. 3d shows the measured constellation diagram transmitted at 6 Mbps. All 64 symbols are clearly separated with a -26 dB error vector magnitude (EVM).

**MIMO Communication**

The multiple beams steered by an AOBS channel in different directions can be modulated by corresponding control signals to transmit different data simultaneously, thereby realizing MIMO optical communication (Fig. 4a). The total communication throughput of each AOBS channel is upper bound by the bandwidth of the IDT, which is 450 MHz in the current device. For multi-beam communication, the available bandwidth is reduced to avoid crosstalk, which requires an inter-beam frequency separation higher than 8 MHz set by beam divergence. The total throughput of the mAOBS array increases proportionally with the number of channels. We demonstrate MIMO optical communication with the scheme as shown in Fig. 4a. Due to the limited number of signal generators, we drove a single AOBS channel with four control signals to generate four beams, each transmitting a different image. Fig. 4b shows the spectrum of the four control signals with frequencies ranging from 2.27 to 2.39 GHz and 40 MHz spacing. Each control signal carries the data stream of an image encoded using 16-QAM return-to-zero coding at 2 Mbps, which helps reduce inter-symbol interference. The transmitted optical signals are received by photodetectors and decoded using digital signal processing (DSP). Fig. 4c displays samples of 20 μs duration of the transmitted signals. The aggregated constellation diagrams after demodulation and corresponding images recovered from the data streams are shown in Fig. 4d. The clearly separated constellation demonstrates the communication fidelity of four beams with a combined data rate of 8 Mbps. If all four channels of the mAOBS array are used, each generating 16 beams and transmitting data at 2 Mbps, the aggregated data rate of the current system will reach 128 Mbps.

**Discussion**

In conclusion, we have demonstrated a multi-beam AOBS array, which monolithically integrates PICs with acousto-optics on a scalable platform. This system, with its compact footprint, high extinction ratio, superior dynamic performance, and broadband modulation capabilities, presents a promising multi-beam steering solution for diverse applications in optical sensing, imaging, and communication. There are several enabling innovations that are worth noting. The use of lithium niobate and the large steering aperture enable AOBS to handle high optical power over a broad spectral range from near-IR to visible, which is significantly advantageous over OPAs and FPSAs that use silicon photonics. The system's high beam quality and broad acoustic bandwidth result in lower than 20 dB inter-beam crosstalk. Future improvement in fabrication quality and device structure are expected to further reduce crosstalk. Additionally, the broadband IDT design provides ample bandwidth for modulating individual beams, making the mAOBS array particularly powerful for applications like LiDAR, quantum gate operations, and MIMO-enabled FSOC. The latter is especially valuable in scenarios such as communication among drone swarms and emergency communication reconstruction, when the ability to communicate distinct information to different receivers is crucial. Further development of this technology will enhance its efficiency and versatility, making it an optimal solution for high-speed, multi-beam imaging, control, and communication in a variety of demanding applications.

## Methods

### Device Fabrication

The mAOBS array is fabricated on x-cut lithium niobate on insulator (LNOI) wafers with 300 nm thick LN layer (from NanoLN Inc.). The optical waveguides are patterned with electron-beam lithography and then etched by inductively coupled plasma reactive ion etcher (ICP-RIE). After stripping the resist, piranha and standard cleaning solution are used to remove resist residue and smooth the surface. The mode converter is patterned with electron-beam lithography using negative resist hydrogen silsesquioxane (HSQ). The IDT was patterned with electron-beam lithography and followed by a lift-off process of 100 nm thick aluminum film.

**Optical measurement setup**

A fiber-coupled laser (Newport TLB-6712) operating at 780 nm is used as the light source in all measurement. An arbitrary waveform generator (Tektronix AWG70001A) is used to generate the control signals for the measurement shown in Fig. 2 and Fig. 3. For multi-beam steering and communication measurement in Fig. 4, multi-tone control signals are generated by a multi-channel FPGA system. The beam profiles in Fig. 2a are captured by a CCD camera (Lumenera Infinity2) at the focal plane of the collimating lens. A beam splitter is placed between the camera and the lens. This enables simultaneous measurement using a fiber to measure beam intensity with a high dynamic range. A fiber is scanned in the focal plane by a 2-axis motorized micropositioning system (MCL Micro-Stage) with closed-loop control to collect the scattered light. The light collected by the fiber is detected by a photodetector (APD130A2). The electrical signal from the photodetector is measured by an oscilloscope (MSO/DPO70000DX).

**Acknowledgments**

This work is supported by the National Science Foundation (Award No. ITE-2134345 and OSI-2326746) and the DARPA MTO SOAR program (Award No. HR0011363032). The authors acknowledge Atom Computing for providing the FPGA system. Part of this work was conducted at the Washington Nanofabrication Facility and Molecular Analysis Facility, a National Nanotechnology Coordinated Infrastructure (NNCI) site at the University of Washington with partial support from the National Science Foundation via award nos. NNCI-2025489. Part of device fabrication was conducted at the Center for Nanophase Materials Sciences (CNMS2022-B-01473), which is a DOE Office of Science User Facility.

# FIGURES

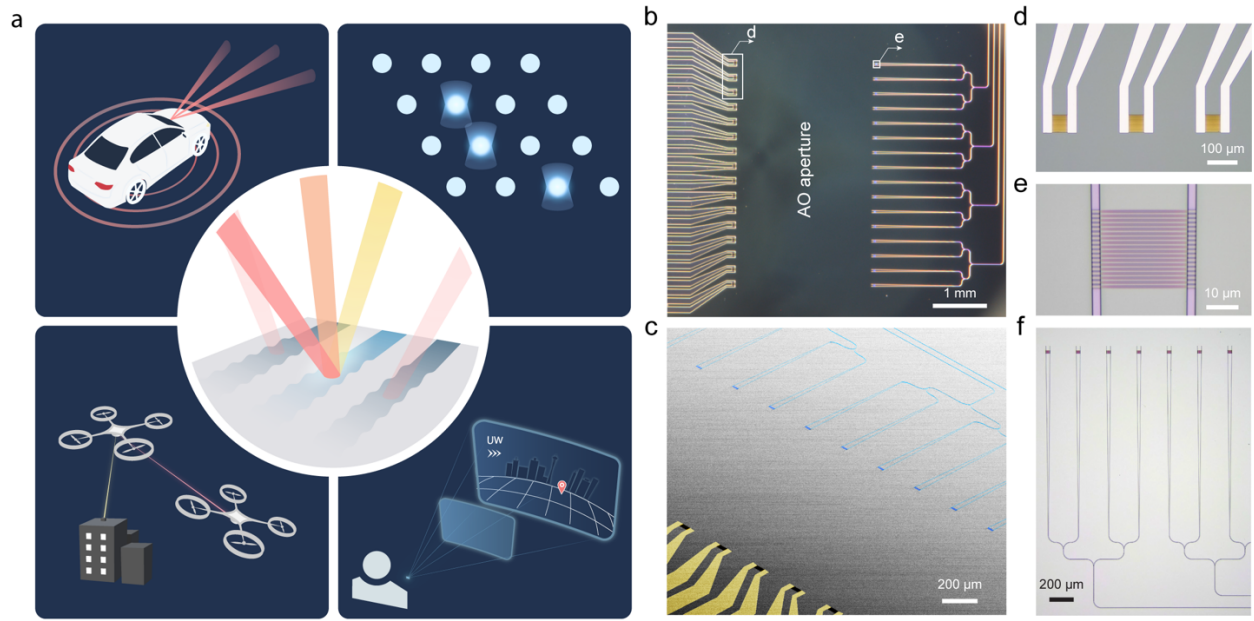

**Fig. 1: Integrated multi-beam acousto-optic steering (mAOBS) array. a**. Optical multi-beam steering finds many important applications ranging from (clockwise from upper left) LiDAR, quantum computing, AR/VR display, and free-space optical communication. **b**. Optical microscope image and **c**. scanning electron microscope (SEM) images of the mAOBS array. The device consists of three main components of **d**. IDT (left white box in b), **e**. mode converter (right white box in b), and **f**. integrated photonic circuits of beam splitters and waveguide tapers.

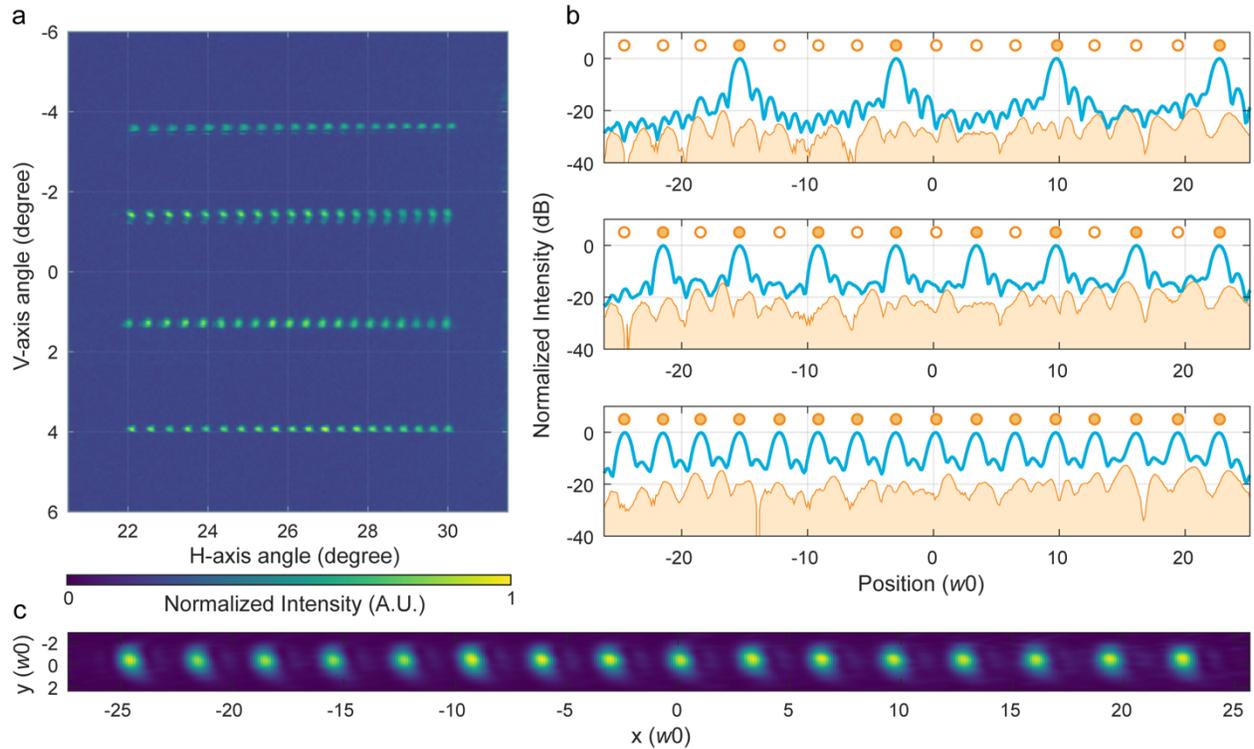

**Fig. 2: Multi-beam steering demonstration and characteristics. a**. Superimposed image 4×21 beams generated by four AOBS channels and captured on the image plane of the collimating lens. The field of view is 8°×8°. Each row is generated by one AOBS channel. Beams in the second row suffers distortion due to a blemish of contamination on the chip surface. **b**. Normalized intensity profile of 4 (top), 8 (middle), and 16 (bottom) beams generated by a single AOBS channel driven by control signals of multiple frequency tones. Cyan curves show the optical intensity at the focal plane while the shaded areas show the background when the control signals are turned off. Circles on the top mark the position and on (solid) or off (hollow) of each site. The x-axis is the beam position normalized to the average beam width $w_0$. **c**. 2D intensity profile of 16 beams generated simultaneously by a single AOBS channel and captured by scanning a fiber on the imaging plane of the collimating lens.

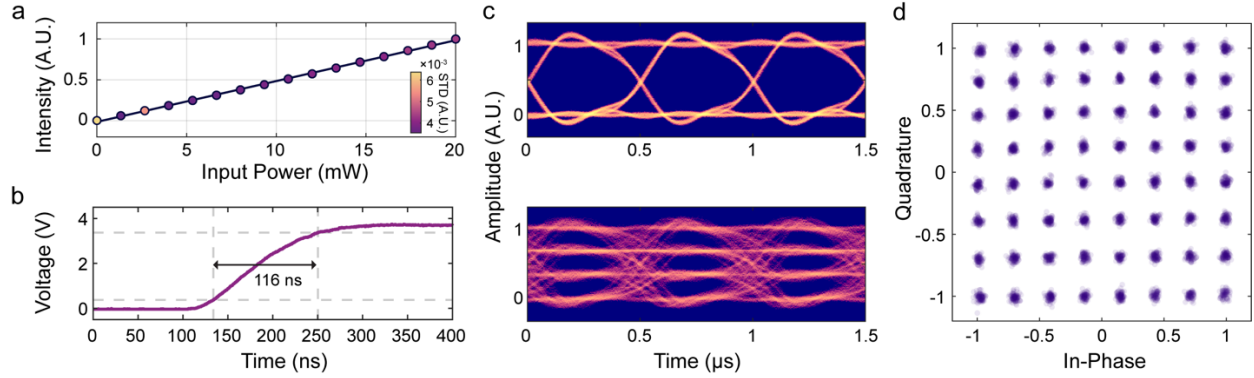

**Fig. 3: Dynamic performance of mAOBS array and application for optical communication.**
**a**. Average beam intensity of a single beam versus. the control signal power (symbols) and the linear fit (line). Each data point is measured for 30,000 times. The normalized standard deviation is indicated by the color in the symbol. **b**. Time trace of the photodetector voltage receiving the beam when the control signal is switched on. The 10-90% rising time is measured to be 116 ns. **c**. Eye diagrams of the AOBS transmitted data of pseudo-random sequence encoded by OOK (upper) at 2 Mbps and PAM4 (lower) at 4 Mbps. **d**. Constellation diagram of the AOBS transmitted data using 64-QAM modulation at 6 Mbps.

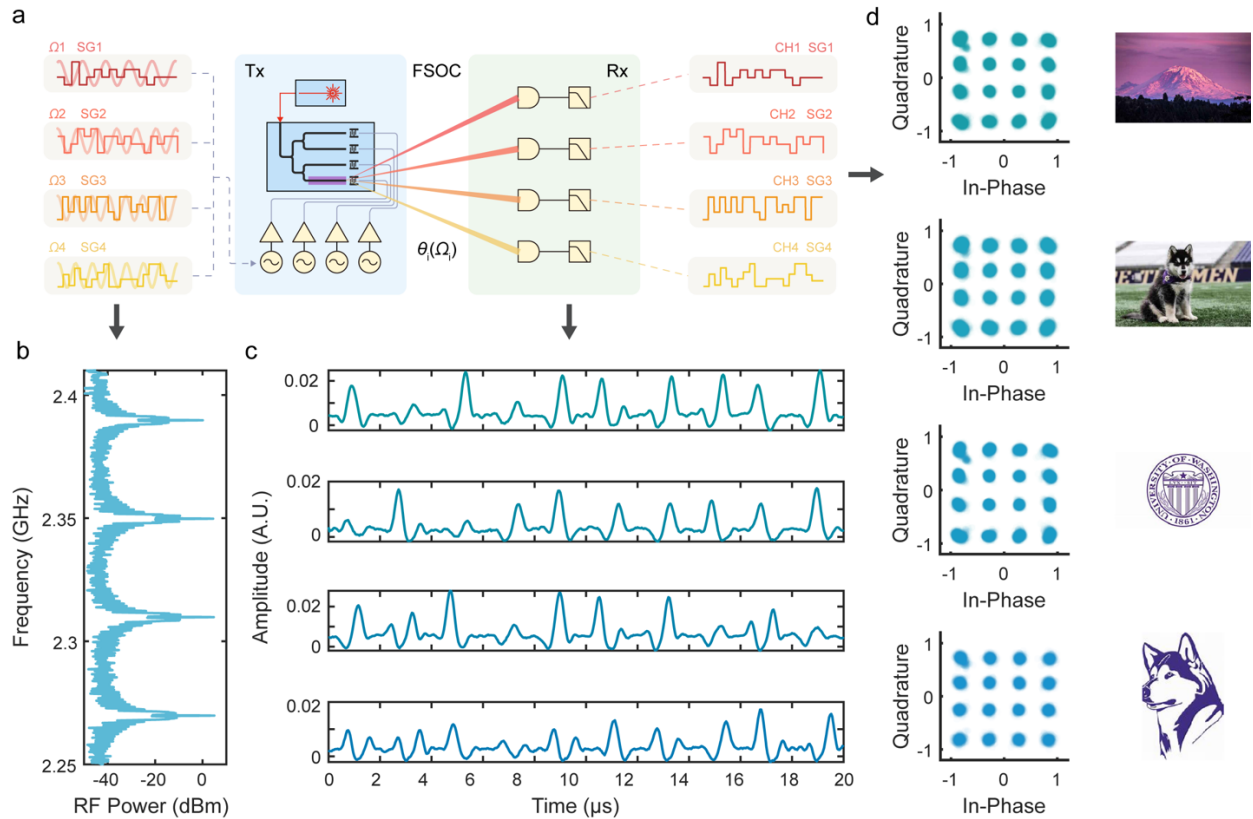

**Fig. 4: Multiple-input-multiple-output (MIMO) enabled free space optical communication using mAOBS array. a**. System schematic of MIMO-enabled FSOC using the AOBS array. A laser source is coupled to the mAOBS array and split to each AOBS channel using PICs. Multiple control signals are used to drive the mAOBS array to steer multiple beams at angles controlled by the control signal frequency. Data is encoded on the beams by amplitude and phase modulating of the control signals. The beams are received by photodetectors, followed by filters and analog-digital converters. The transmitted signals are demodulated by digital signal processing to decode the data. **b**. Spectrum of the control signals driving one AOBS channel with four different central frequencies from 2.27 to 2.39 GHz with 40 MHz spacing. Each control signal carries the data stream of an image encoded in 16-QAM at a data rate of 2 Mbps. **c**. Sections of photodetector signals when receiving four beams steered to different positions. **d.** The constellation diagrams (left) and the corresponding recovered images (right) transmitted by the mAOBS array.